\documentstyle[12pt]{article}
\begin{document}
 
\title {\bf An approach to the problem of generating
irreducible polynomials over the finite field $GF(2)$ and its relationship 
with the problem of periodicity on the space of binary sequences}
\author{Ricardo L\'{o}pez-Ruiz  \\
                                   \\
\small Department of Computer Science and BIFI, \\
\small Facultad de Ciencias - Edificio B, \\
\small Universidad de Zaragoza, \\
\small 50009 - Zaragoza (Spain)}
\date{ }

\maketitle
\baselineskip 6mm
 
\begin{center} {\bf Abstract} \end{center}
A method for generating irreducible polynomials of degree $n$ over the finite field $GF(2)$
is proposed. The irreducible polynomials are found by solving a system of equations that 
brings the information on the internal properties of the splitting field $GF(2^n)$ . 
Also, the choice of a primitive normal basis allows us to build up a natural 
representation of $GF(2^n)$ in the space of $n$-binary sequences.
Illustrative examples are given for the lowest orders.
\newline


\newpage

\section{Introduction}

The problem of determining the irreducible polynomials over a 
finite field of order $q$, $GF(q)$, is a task,
from a qualitative mathematical point of view, similar 
to that of finding the prime numbers in the set of integer numbers \cite{hardy,riesel}. 
Such polynomials are necessary in numerous applications.
For instance, the generation of algebraic error-correcting codes and cryptography  
\cite{berlekamp}  or the construction of fast transforms are based on finite 
field properties \cite{reed} that require algorithms to obtain 
the irreducible polynomials $f_n(x)$ of degree $n$ over a particular finite field.

Tables of irreducible polynomials are found in the literature \cite{marsch}. 
These can be built up by different methods \cite{golomb,shoup},
including a close analog to the 'sieve' method used in making tables of prime numbers. 
In this work a method to generate the irreducible polynomials
of degree $n$ over the field $GF(2)$ is proposed. The strategy comprises 
the construction of a representation of 
the splitting field $GF(2^n)$  over the set of binary $n$-digits sequences.
Each periodic orbit and its $m$ permutations, with $m$ a divisor of $n$, identify 
a irreducible polynomial of degree $m$ and its $m$ roots on $GF(2^n)$. 

In section $II$ the connection between the field $GF(2^n)$ and the 
space of $n$-binary sequences, ${\cal B}_n$, is identified.
The two internal operations over ${\cal B}_n$
necessary in order to obtain a particular representation 
of $GF(2^n)$ are defined in section $III$. 
The method to find the irreducible polynomials $f_n(x)$ over $GF(2)$ is 
disclosed in section $IV$. Section $V$ summarizes the
general scheme to be followed in order to obtain $f_n(x)$ for all $n$.
Two illustrative examples are given in section $VI$.  
Last section is concerned with our conclusions.

\section{$GF(2^n)$ and ${\cal B}_n$}

{\bf Notation}: Let us call ${\cal B}_n$ the space of binary sequences 
of $n$ digits. ${\cal B}_n$ has $2^n$ elements. Each element $O\in{\cal B}_n$
is said to be an orbit. If $O$ is given by the binary sequence
($\alpha_{n-1}\alpha_{n-2}\cdots\alpha_1\alpha_0$), ($\alpha_i = 0\;\; or\;\; 1$),
the left shifted orbit $O'$ defined by the sequence  
($\alpha_{n-2}\alpha_{n-3}\cdots\alpha_1\alpha_0\alpha_{n-1}$)
is represented by $shift(O)$. If $m$ is the smallest integer
such that $O=shift^m(O)=shift~(~\stackrel{m}{\cdots}~(~shift~(~O)~)~)$ then 
the orbit $O$ is said to be of {\it period $m$} or $m$-periodic,
and it is represented by $O_m$.
It is obvious that if $O_m\in {\cal B}_n$ then $m$ is a divisor of $n$, 
$m\mid n$. The set $[O_m]$ formed by the $m$ different shifts of 
an $m$-periodic orbit, $O_m$, is called an orbital of period $m$:
$[O_m]=\{O_m,shift(O_m),shift^2(O_m),\cdots,shift^{m-1}(O_m)\}$.
As an example, $[O_3]=[100]=\{100,010,001\}$ or $[O_2]=[0101]=\{0101,1010\}\equiv [01]=\{01,10\}$.

{\bf Fundamental Idea}: 
Firstly, let us note that the number $N_n$ of irreducible polynomials (I.P.)
of degree $n$ over $GF(2)$,
is equal to the number of orbitals $[O_n]$ of period $n$ 
(table~\ref{tabla1},\cite{lopez,mil}):
\begin{equation} 
N_n=\frac{2^n-\sum_{i=1}^{k}m_iN_{m_i}}{n},
\end{equation}
where $\{m_1,m_2,\ldots,m_k\}$ are the integer divisors of $n$
less than $n$.  \newline
The idea is to build up a representation of $GF(2^n)$ over ${\cal B}_n$,
with the constraint that each orbital $[O_m]$ contained in ${\cal B}_n$ has
an associated I.P. , $f_m(x)$, of degree $m$ over $GF(2)$. Moreover, 
the $m$ orbits $O_m$ that expand the orbital $[O_m]$ are the $m$ roots 
of  $f_m(x)$ in $GF(2^n)$.
Thus, two internal operations $(+,*)$
are defined over ${\cal B}_n$ in such a way that $({\cal B}_n,+,*)$  
be the splitting field for all I.P. $f_m(x)$ over $GF(2)$ , with $m\mid n$.

\section{Field Construction in ${\cal B}_n$}

{\bf Choice of a Basis}: 
If we consider $GF(2^n)$ a vectorial space over $GF(2)$,
a {\it primitive normal basis},
$B_{\sigma n}$=\{$\sigma_0$, $\sigma_1$,..., $\sigma_{n-1}$\},
can be found in $GF(2^n)$ \cite{lidl,lenstra1}.
$B_{\sigma n}$ to be {\it normal} means that
$\sigma_i^2=\sigma_{(i+1)\bmod n}$. $B_{\sigma n}$ to be {\it primitive} 
means that $\sigma_0$ generate $GF(2^n)^*=GF(2^n)-\{0\}$.
Therefore, if we denote by $(+,*)$ the two internal operations in $GF(2^n)$, 
each element $a\in GF(2^n)^*$ has a unique representation 
as {\it addition} ($+$) or as {\it product} ($*$) of the $B_{\sigma n}$ elements 
(tables~\ref{tabla2},~\ref{tabla3},~\ref{tabla4},~\ref{tabla5},~\ref{tabla6}): \newline
\hspace{1cm} (i) \underline{\underline{{\it $+$-representation}}}: 
each element $a\in GF(2^n)$ can be expanded
in a linear combination of $B_{n\sigma}$, $a=\sum_{i=0}^{n-1}\alpha_i\sigma_i$ with 
$\alpha_i\in GF(2)\equiv\{0,1\}$. The binary sequence 
$a_+=(\alpha_{n-1}\alpha_{n-2}\cdots\alpha_1\alpha_0)\in {\cal B}_n$
is the $+$-representation of $a$. \newline
\hspace{1cm} (ii) \underline{\underline{{\it $*$-representation}}}: 
each element $a\in GF(2^n)^*$ is generated 
by $\sigma_0$, consequently $a=\sigma_0^k$ with $k\in\{1,2,\cdots,2^n-1\}$. 
If we take the binary expression of the exponent,
$k=\sum_{i=0}^{n-1}\beta_i2^i$, the binary sequence 
$a_*=(\beta_{n-1}\beta_{n-2}\cdots\beta_1\beta_0)\in {\cal B}_n$
is the $*$-representation of $a$. It can also be reformulated as
$a=\Pi_{i=0}^{n-1}\sigma_i^{\beta_i}$ where $\sigma_i^0=1$. It is
obvious that $(a^2)_*=shift(a_*)$.

\underline{Example} (table~\ref{tabla5}): 
If $B_{\sigma 4}$=\{$\sigma_0$, $\sigma_1$, $\sigma_2$, $\sigma_3$\}
is a primitive normal basis, 
the following expansions are obtained for some elements of $GF(2^4)$:
\begin{eqnarray*}
a = \sigma_0^3 = \sigma_0+\sigma_1+\sigma_3 & \rightarrow & a_+ = 1011, \\
a = \sigma_0^3 = \sigma_0*\sigma_1 & \rightarrow & a_* = 0011, \\
& & \\
b = \sigma_0^4 = \sigma_2 & \rightarrow & b_+ = 0010, \\
b = \sigma_0^4 = \sigma_2 & \rightarrow & b_* = 0010, \\
 & & \\
c = \sigma_0^7 = \sigma_2+\sigma_3 & \rightarrow & c_+ = 1100, \\
c = \sigma_0^7 = \sigma_0*\sigma_1*\sigma_2 & \rightarrow & c_* = 0111. 
\end{eqnarray*}
The particular elements, $a=\sigma_0^{2^i}=\sigma_i$ and 
$a=\sigma_0^{2^n-1}=\sigma_{n-1}*\sigma_{n-2}*\cdots*\sigma_1*\sigma_0=
\sigma_{n-1}+\sigma_{n-2}+\cdots+\sigma_1+\sigma_0=1$, verify $a_+=a_*$.
In this case,  $(00\cdots 010\cdots 00)$ and $(11\cdots 11\cdots 11)$, respectively.

{\bf Internal operations in ${\cal B}_n$}: we explain now how to perform  
the two internal operations $(+,*)$ in ${\cal B}_n$. 

\underline{\underline{\it $+$-operation}}: the property $\sigma_i+\sigma_i=0$ 
(the characteristic of $GF(2^n)$ is $2$) imposes the algorithm for performing this operation.
Given $a,b\in GF(2^n)$ the element $a+b$ is obtained by summing bit to bit the binary
sequences of $a$ and $b$ ($\bmod 2$) in the $+$-representation. If we have
\begin{eqnarray*}
a_+ & = & \alpha_{n-1}\alpha_{n-2}\cdots\alpha_1\alpha_0, \\
b_+ & = & \beta_{n-1}\beta_{n-2}\cdots\beta_1\beta_0, \\
(a+b)_+ & = & \gamma_{n-1}\gamma_{n-2}\cdots\gamma_1\gamma_0,
\end{eqnarray*}
the addition $+$ presents the form:
\begin{eqnarray*}
(a+b)_+ & = & a_+ + b_+ \bmod 2,  \\
\gamma_i & = & (\alpha_i + \beta_i) \bmod 2. 
\end{eqnarray*}
$({\cal B}_n, +)$ is a representation of the abelian additive group defined over $GF(2^n)$,
where the neutral element is $0\equiv (00\cdots 00)$ and the inverse element 
$\overline a$ of $a$ is the element
itself, $a+a=0$, then $\overline{a}=a$. \newline
\underline{Example} (table~\ref{tabla5}):
\begin{eqnarray*}
(\sigma_0^3)_+ & \rightarrow & 1011, \\
(\sigma_0^4)_+ & \rightarrow & \underline{0010}, \\
(\sigma_0^3+\sigma_0^4)_+ & \rightarrow & 1001 =\sigma_0^{14}.
\end{eqnarray*}

\underline{\underline{\it $*$-operation}}: the property $\sigma_i^2=\sigma_{(i+1) \bmod n}$ gives 
the clue to perform this operation.
Given $a,b\in GF(2^n)$ the element $a*b$ is obtained after 
performing the addition with carrier ($\bmod 2$) of the binary
sequences of $a$ and $b$ in the $*$-representation. If we have
\begin{eqnarray*}
a_* & = & \alpha_{n-1}\alpha_{n-2}\cdots\alpha_1\alpha_0, \\
b_* & = & \beta_{n-1}\beta_{n-2}\cdots\beta_1\beta_0, \\
(a+b)_* & = & \gamma_{n-1}\gamma_{n-2}\cdots\gamma_1\gamma_0, 
\end{eqnarray*}
then the multiplication $*$ presents the form:
\begin{eqnarray*}
(a*b)_* \equiv a_* * b_* & = & (a_* + b_*\; with\;\;carrier)\bmod 2,\;\;  \\
\gamma_i & = & (\alpha_i + \beta_i  + carrier_{i})\bmod 2.
\end{eqnarray*}
To start we suppose $carrier_0=0$. If $\alpha_{n-1}+\beta_{n-1}+carrier_{n-1}>1$ then the operation is restarted
with $carrier_0=1$. 
$({\cal B}_n, *)$ is a representation of the abelian  multiplicative group defined 
over $GF(2^n)^*$, where the neutral element is 
$1\equiv (11\cdots 11)$ and the inverse element $a^{-1}$ of $a$ is represented by 
the complementary binary sequence, $(a^{-1})_*=cp(a_*)$
given by $digit_i((a^{-1})_*)=(digit_i(a_*)+1\;\;\bmod 2$).
This is a consequence of the fact that if $a=\sigma_0^k$ then $a^{-1}=\sigma_0^{(2^n-1)-k}$. 
The multiplication by zero is $(a*0)_*=a_**(00\cdots 00)=(00\cdots 00)\equiv 0$ by definition.
It can also be seen that if $(a*b)_*=0$, then only one of the two sequences are null, 
$a=0$ or $b=0$. \newline
\underline{Example}:
\begin{eqnarray*}
carrier_0=0, & & \\
(\sigma_0^{11})_* & \rightarrow & 1011, \\
(\sigma_0^4)_* & \rightarrow & \underline{0010}, \\
(\sigma_0^{11}*\sigma_0^4)_*& \rightarrow & 1101 =\sigma_0^{13}.
\end{eqnarray*}
\begin{eqnarray*}
carrier_0=1, & & \\
(\sigma_0^{11})_* & \rightarrow & 1011, \\
(1)_* & \rightarrow & \underline{1111}, \\
(\sigma_0^{11}*1)_*& \rightarrow & 1011 =\sigma_0^{11}.
\end{eqnarray*}
\begin{eqnarray*}
a=\sigma_0^{11} & \rightarrow & a_* = 1011, \\
a^{-1}=\sigma_0^{4} & \rightarrow & (a^{-1})_*=0100.
\end{eqnarray*}
\underline{\underline{{\it Connection between $+$ and $*$}}}: 
the central question in order to build up
a representation of the field $GF(2^n)$ over ${\cal B}_n$
is to find the relationship between $a_+$ and $a_*$.
This construction can be performed if that relation 
is known for the elements $\sigma_i*\sigma_j$ because
all the other elements can be reduced to products of $\sigma_i*\sigma_j$ type.
Thus, if we write
\begin{eqnarray*}
\sigma_i*\sigma_j & = & \sum_{l=0}^{n-1} e_{ijl}\sigma_l,
\end{eqnarray*}
there are $n^3$ constants $e_{ijl}$ ($=0,1$) defining the field $GF(2^n)$ completely.
By construction  $\sigma_i*\sigma_i=\sigma_{i+1}$, then $e_{iil}=\delta_{(i+1)l}$.
By the commutative property, $e_{ijl}=e_{jil}$.
Also, $(\sigma_i*\sigma_j)^2=\sigma_{i+1}*\sigma_{j+1}$, then  
$shift((\sigma_i*\sigma_j)_*)=(\sigma_{i+1}*\sigma_{j+1})_*$.
If we impose this property for the $+$-representation, that is,
$shift((\sigma_i*\sigma_j)_+)=(\sigma_{i+1}*\sigma_{j+1})_+$, then
$e_{(i+1)(j+1)(l+1)}=e_{ijl}$.
Thus, only $(n-1)/2$ products $\sigma_i*\sigma_j$ when $n$ is odd, and
$n/2$ products  $\sigma_i*\sigma_j$ when $n$ is even, must be known in 
order to determine the field completely. That is, a total number 
of $n^2/2$ or $(n^2-n)/2$ constants $e_{ijl}$, depending 
on the parity of $n$, defines the field. These constants cannot be freely chosen 
as the distributive property must be verified:
\begin{eqnarray*}
a*(b+c) & = & a*b + a*c, \\
\left( a_* * (b_+ + c_+)_*\right)_+ & = & (a_* * b_*)_+ + (a_* * c_*)_+.
\end{eqnarray*}
All the constants $e_{ijl}$ stay fixed by these conditions in some cases and 
the natural construction $({\cal B}_n,+,*)$ is unique.
In general, this is not the case and there exist different constructions 
of $GF(2^n)$ in ${\cal B}_n$.

\section{Construction of the Irreducible Polynomials}

Let us recall that $GF(2^n)$ is the splitting field of all irreducible polynomials
$f_m(x)$ over $GF(2)$ of degree $m\leq n$ and $m\mid n$. 
If $a\in GF(2^n)$ is a root of $f_m(x)$, the $m$ roots of this irreducible
polynomial are $\{a,a^2,a^{2^2},\cdots,a^{2^{(m-1)}}\}$.
Therefore, the $m$ orbits that integrate an orbital $[O_m]\in {\cal B}_n$
are the $m$ roots of the irreducible polynomial $f_m(x)$ in the representation 
$({\cal B}_n,+,*)$ of $GF(2^n)$, as explained in the last section. 
For instance, in the case $n=4$ (table~\ref{tabla5}), the elements 
$\{\sigma_0\sigma_1,\sigma_1\sigma_2,\sigma_2\sigma_3,\sigma_3\sigma_0\}$ are the roots
of the irreducible polynomial $x^4+x^3+x^2+x+1$. The objetive of this section is to establish
a method to find the irreducible polynomials over $GF(2)$ working out this property. 

{\bf Trace of an orbital}: The trace of an orbital $[O_m]$, $Tr(O_m)$, 
is defined as the trace of one of its elements:
$\{a,a^2,a^{2^2},\cdots,a^{2^{(m-1)}}\}$ with $a\in GF(2^n)$, that is,
\[ Tr(O_m) = a+a^2+a^{2^2}+\cdots +a^{2^{(m-1)}}. \]
The trace is an element of $GF(2)$, in this case $0$ or $1$.
The number of traces $N_T^{(n)}$ in $GF(2^n)$ is equal to the number of 
orbitals $[O_m]$ in ${\cal B}_n$,
\begin{displaymath}
 N_T^{(n)}=\sum_m N_m \;\;\; with \;\; m\mid n.
\label{eqNT}
\end{displaymath} 
Let us call $N_{T1}^{(n)}$ the number of traces whose value is $1$ 
and $N_{T0}^{(n)}$ the number of those whose value is $0$.
Hence, $N_T^{(n)} = N_{T0}^{(n)} + N_{T1}^{(n)}$.\newline
The important point for our proposal is that the coefficients of an 
I.P.  over $GF(2)$ can be obtained by adding ($\bmod 2$) the traces 
of different orbitals of the field $GF(2^n)$.
Therefore the I.P. can be identified
if the traces are previously calculated. 

\underline {Example} (table~\ref{tabla5}): (i) the I.P. associated to the orbital
$\{\sigma_0\sigma_1,\sigma_1\sigma_2,\sigma_2\sigma_3,\sigma_3\sigma_0\}$ is 
a polynomial of degree $4$, $x^4 + d x^3 + e x^2 + f x + 1$. Its coefficients
are functions of the traces:
\begin{eqnarray*}
d = & \sigma_0\sigma_1+\sigma_1\sigma_2+\sigma_2\sigma_3+\sigma_3\sigma_0
 & = Tr(\sigma_0\sigma_1), \\
e = & \sigma_0\sigma_1+\sigma_1\sigma_2+\sigma_2\sigma_3+\sigma_3\sigma_0
+\sigma_0\sigma_1\sigma_2\sigma_3+\sigma_0\sigma_1\sigma_2\sigma_3
 & = Tr(\sigma_0\sigma_1), \\
f = &  \sigma_0\sigma_1+\sigma_1\sigma_2+\sigma_2\sigma_3+\sigma_3\sigma_0
 & = Tr(\sigma_0\sigma_1).
\end{eqnarray*}
(ii) the I.P. associated to the orbital $\{\sigma_0, \sigma_1, \sigma_2, \sigma_3\}$
is also of degree $4$, $x^4 + a x^3 + b x^2 + c x + 1$, where:
\begin{eqnarray*}
a = & \sigma_0+\sigma_1+\sigma_2+\sigma_3 & = Tr(\sigma_0), \\
b = & \sigma_0\sigma_1+\sigma_1\sigma_2+\sigma_2\sigma_3+\sigma_3\sigma_0+
\sigma_0\sigma_2+\sigma_1\sigma_3
 & = Tr(\sigma_0\sigma_1) + Tr(\sigma_0\sigma_2), \\
f = & \sigma_0\sigma_1\sigma_2+\sigma_1\sigma_2\sigma_3+
\sigma_2\sigma_3\sigma_0+\sigma_3\sigma_0\sigma_1 & = Tr(\sigma_0\sigma_1\sigma_2).
\end{eqnarray*}

{\bf Newton's Formulas}: the {\it fundamental theorem on symmetrical polynomials}
can guide us in finding the different traces in the field $GF(2^n)$.
This theorem can be expressed as follows \cite{edwards}:
Let $r_1$, $r_2$, $\cdots$, $r_n$ be different variables and let us call 
$\theta_1$, $\cdots$, $\theta_n$ the following symmetrical polynomials in those
variables:
\begin{eqnarray*}
\theta_1 & = & r_1 + r_2 + \cdots + r_n, \\
\theta_2 & = & r_1 r_2 + r_1r_3 + \cdots + r_{n-1}r_n, \\
\theta_3 & = & r_1 r_2r_3 + r_1r_2r_4 + \cdots + r_{n-2}r_{n-1}r_n, \\
  & \vdots &  \\
\theta_n & = & r_1 r_2 r_3 \cdots r_n .
\end{eqnarray*}
Then every symmetrical polynomial in the variables
\{$r_1$, $r_2$, $\cdots$, $r_n$\}  and with coefficients over a field $F$ 
can be expressed as a polynomial in the new variables
$\{\theta_1,\theta_2,\cdots,\theta_n\}$ with coefficients in $F$.

In the particular case of the following symmetrical polynomials $s_k$:
\[ s_k = r_1^k + r_2^k +\cdots + r_n^k \;\;\;\; k = 1,2,3,\cdots \]
the expansion in the variables $\theta_k$ is given by the {\it Newton's formulas} \cite{newton}:
\begin{eqnarray}
s_1 - \theta_1 & = & 0, \nonumber \\
s_2 - s_1\theta_1 + 2\theta_2 & = & 0,  \nonumber \\
s_3 - s_2\theta_1 + s_1\theta_2 + 3\theta_3 & = & 0,  \label{eqsk} \\
\vdots & & \nonumber \\
s_k - s_{k-1}\theta_1 + s_{k-2}\theta_2 - \cdots +
 (-1)^{k-1}s_1\theta_{k-1} + (-1)^k(k)\theta_k & = & 0, \nonumber 
\end{eqnarray}
where $\theta_j=0$ for $j>n$ and  the symbol $(k)$ is interpreted as
$(k)=1+1+\stackrel{k}{\cdots}+1$. \par
In our problem, for instance, if we take $r_i=\sigma_{i-1}$ the expressions of
$s_k$ and $\theta_k$ are functions of the $GF(2^n)$ traces. Then 
the Newton's formulas give us a system of equations for the different traces of $GF(2^n)$.
By solving this system it is possible build up the I.P. of degree $n$ over $GF(2)$.

{\bf Finding the Traces}: If the elements $\{\sigma_0,\sigma_1,\cdots,\sigma_{n-1}\}
 \in GF(2^n)$ are taken as the variables \{$r_1$, $r_2$, $\cdots$, $r_n$\} in the 
equations~(\ref{eqsk}), a system of equations for the different traces is obtained.
The total number of unknowns (traces) is equal to the total number of orbitals,
$N_T^{(n)}$, in  $GF(2^n)$ (eq.~\ref{eqNT}). By construction:
\begin{eqnarray*}
Tr(0) & = & 0, \\
Tr(\sigma_0) & = & 1, \\
Tr(\sigma_0\sigma_1\cdots\sigma_{n-1}) & = & 1.
\end{eqnarray*}
For large $n$ the number of orbitals is of the same order as
\[ N_T^{(n)}\sim \frac{2^n}{n}. \] 
Noting that $r_i^{2^n}=\sigma_i^{2^n}=\sigma_i=r_i$ we will have $2^n$ equations. Therefore, 
a system of $2^n$ equations with $2^n/n$ unknowns (many equations are
interdependent) should be solved. Only those solutions verifying 
that the number of traces  with value $1$ is equal to $N_{T1}^{(n)}$ are accepted.
If one of these solutions is chosen, then the I.P. are directly calculated.

{\bf Note}: Let us recall a property  that allows us 
to generate the I.P. by pairs. \newline
Two polynomials, $f(x)$ and $g(x)$, are reciprocal if they can be expressed in the form:
\begin{eqnarray*}
f(x) & = & a_nx^n + a_{n-1} x^{n-1} + \cdots + a_1 x + a_0,  \\
g(x) & = & a_0x^n + a_1 x^{n-1} + \cdots + a_{n-1} x + a_n,  
\end{eqnarray*}
or equivalently, $g(x)=x^nf(x^{-1})$. This means that the roots of $g(x)$
are the inverse roots of $f(x)$. Consequently, if $f_m(x)$ is an I.P. of degree $m$, 
the reciprocal polynomial $g_m(x)$ is also an I.P. of degree $m$. 
This property can be used as another useful tool to facilitate the finding
of all irreducible polynomials. It is necessary to remark at this point
that both polynomials can be the same, that is, $f_m(x)=g_m(x)$.

\section{General Methodology}

Summarizing, the steps to follow for building up a natural representation of the field $GF(2^n)$ 
in ${\cal B}_n$ and for finding all I.P. of degree $n$ over $GF(2)$ are the following:

{\bf (1)} All the independent $n$-binary sequences (orbitals $[O_m]$ with $m\mid n$) in ${\cal B}_n$
are found (see appendix). Its total number, $N_{T}^{(n)}$, is the total number of I.P.
over $GF(2)$. The splitting field is $GF(2^n)$. Let us observe that each orbital $[O_m]$
has associated an I.P. $f_m(x)$ of degree $m$. The roots of $f_m(x)$ in  $({\cal B}_n,+,*)$ are 
the $m$ roots given by the $m$ orbits of the orbital $[O_m]$. 
It also is possible to evaluate $N_{T1}^{(n)}$ by calculating directly
the trace for each orbit $O_m$,
\[ Tr(O_m) = O_m + shift(O_m) + shift^2(O_m) + \cdots + shift^{m-1}(O_m), \]
where $shift(O_m)$ is the left shift of the orbit $O_m$. If $\#_1(O_m)$ 
represents the number of $1$ in the orbit $O_m$ multiplied by $m/n$ then
it can be deduced that
\begin{eqnarray*}
Tr(O_m) = 1 & if & \#_1(O_m)\;\; is\;\; odd, \\
Tr(O_m) = 0 & if & \#_1(O_m)\;\; is\;\; even.
\end{eqnarray*}
For instance,
\begin{eqnarray*}
\#_1(0011)=2\cdot\frac{4}{4}=2 & \Rightarrow & Tr(0011) = 0, \\
\#_1(0111)=3\cdot\frac{4}{4}=3 & \Rightarrow & Tr(0111) = 1, \\
\#_1(0101)=2\cdot\frac{2}{4}=1 & \Rightarrow & Tr(0101) = 1, \\
\#_1(1111)=4\cdot\frac{1}{4}=1 & \Rightarrow & Tr(1111) = 1. 
\end{eqnarray*}

{\bf (2)} To calculate the trace associated to each element in $GF(2^n)$,
the $2^n$ non-independent equations deriving from the Newton's formulas
are established. There are $N_{T}^{(n)}\sim 2^n/n$  unknowns.

{\bf (3)} A solution of the former equation system is worked out. 
This  solution must be compatible with the constraints: (i) $Tr(0)=0$,
$Tr(\sigma_0)=1$ and $Tr(\sigma_0\sigma_1\cdots\sigma_{n-1})=1$; (ii) the number 
of traces with value $1$ must be equal to $N_{T1}^{(n)}$
(calculated in point (1)). 
The different possible solutions are related with the different possible 
constructions of $GF(2^n)$ in ${\cal B}_n$.

{\bf (4)} The irreducible polynomials are explicitely found in a 
deterministic way from a particular solution obtained at point (3).

{\bf (5)} The construction of $({\cal B}_n, +, *)$ is performed
from the I.P. calculated in (4). This is a natural representation of $GF(2^n)$.

{\bf (6)} \underline{Note}:  As it can be seen in the tables below,  
an only representation of $GF(2^n)$ in ${\cal B}_n$ is possible when $\{n=1,2,3,4\}$.

\section{Examples}

In this section the process formerly explained is applied for the cases $n=4,5$.

 {\bf n=4}:
Let us suppose $GF(2^4)$  expanded by the primitive normal
basis $B_{\sigma 4}=\{\sigma_0,\sigma_1,\sigma_2,\sigma_3\}$.
Then $\sigma_{i}^2=\sigma_{(i+1)\bmod 2}$. The field's elements
$\{0,\sigma_0,\sigma_0^2,\sigma_0^3,\cdots,\sigma_0^{14},
\sigma_0^{15}=1\}$ are generated by $\sigma_0$.
If an element $a\in GF(2^4)$ verifies $a^{2^m}=a$, the $m$ powers of
$a$, $\{a,a^2,a^{2^2},\cdots,a^{2^{m-1}}\}$,
are the $m$ roots of an I.P. of degree $m$ over $GF(2)$.
Obviously, $m$ is a divisor of $4$. 
The different representative elements of this kind in $GF(2^4)$ are:
\begin{center}
\begin{tabular}{|l|l|} \hline
 $a\in GF(2^4)$ & $m$                \\ \hline\hline 
 $0 $           & $1$           \\
 $\sigma_0$     & $4$ 	   \\
 $\sigma_0\sigma_1$  & $4$     \\
 $\sigma_0\sigma_2$  & $2$     \\
 $\sigma_0\sigma_1\sigma_2$  & $4$         \\
 $\sigma_0\sigma_1\sigma_2\sigma_3$ & $1$  \\ \hline
\end{tabular}
\end{center}
The  different orbitals $[O_m]$ of period $m$, with $m\mid 4$,
in ${\cal B}_4$ are given by the representative orbits:
\begin{center}
\begin{tabular}{|l|l|l|} \hline
$[O_m]$  & $Tr(O_m)$ & m                \\ \hline\hline 
$0000$ &      0     & $1$   \\
$0001$ &      1	    & $4$  \\
$0011$ &      0     & $4$  \\  
$0101$ &      1     & $2$  \\
$0111$ &      1     & $4$  \\
$1111$ &      1     & $1$  \\   \hline
\end{tabular}
\end{center}
Thus, $N_T^{(4)}=N_1+N_2+N_3=6$ and $N_{T1}^{(4)}=4$ (see Table 1). It can also be
observed that the orbit $\{\sigma_0\sigma_2,\sigma_1\sigma_3\}$ is of period $2$, then its
binary $+$-representation is $\{0101,1010\}$ 
and by direct calculation $Tr(\sigma_0\sigma_2)=1$. \newline
If $\{\sigma_0,\sigma_1,\sigma_2,\sigma_3\}$ are taken as the roots of an I.P. of degree 4,
the equations derived from Newton's formulas (where $\theta_k=0$ for $k>4$) for the different traces
can be established. The variables $\theta_k$ and $s_k$ are expressed as follows: 
\begin{eqnarray*}
\theta_1  & = & \sigma_0 + \sigma_1 + \sigma_2 + \sigma_3  \\
	& = & Tr(\sigma_0) = 1, \\
\theta_2  & = & \sigma_0\sigma_1 + \sigma_1\sigma_2 + \sigma_2\sigma_3 + \sigma_3\sigma_0 +
	\sigma_0\sigma_2 + \sigma_1\sigma_3 \\
	& = & Tr(\sigma_0\sigma_1) + Tr(\sigma_0\sigma_2), \\
\theta_3 & = & \sigma_0\sigma_1\sigma_2 + \sigma_1\sigma_2\sigma_3 + 
	\sigma_2\sigma_3\sigma_0 +  \sigma_3\sigma_0\sigma_1 \\
	& = & Tr(\sigma_0\sigma_1\sigma_2), \\
\theta_4 & = & \sigma_0\sigma_1\sigma_2\sigma_3 \\
	& = & Tr(\sigma_0\sigma_1\sigma_2\sigma_3) = 1,\\
s_k	& = & \sigma_0^k + \sigma_1^k + \sigma_2^k + \sigma_3^k \\
	& = & \sum_i Tr(\sigma^i) \;\; with\;\; k=1,2,\cdots,15.
\end{eqnarray*}
The independent Newton's formulas from the system~(\ref{eqsk}) are:
\begin{eqnarray*}
Tr(\sigma_0\sigma_2) & = & 1, \\
Tr(\sigma_0\sigma_2) + Tr(\sigma_0\sigma_1\sigma_2) & = & 1, \\
Tr(\sigma_0\sigma_1) & = & 1. 
\end{eqnarray*}
The solution is unique:
\begin{eqnarray*}
Tr(\sigma_0) & = & 1, \\
Tr(\sigma_0\sigma_1) & = & 1, \\
Tr(\sigma_0\sigma_2) & = & 1,\\
Tr(\sigma_0\sigma_1\sigma_2) & = & 0, \\
Tr(\sigma_0\sigma_1\sigma_2\sigma_3) & = & 1. \\
\end{eqnarray*}
Finally, the I.P. are:
\begin{eqnarray*}
\; [O_4]\equiv\{\sigma_0\} \rightarrow & 
	x^4 + a x^3 + b x^2 + c x + 1 & \rightarrow x^4 + x^3 + 1, \\
\; [O_4]\equiv\{\sigma_0^{-1}=\sigma_0\sigma_1\sigma_2\} \rightarrow & 
	x^4 + c x^3 + b x^2 + a x + 1 & \rightarrow x^4 + x + 1,  \\
 & a = Tr(\sigma_0) = 1, & \\
 & b = Tr(\sigma_0\sigma_1) + Tr(\sigma_0\sigma_2) = 0, & \\
 & c = Tr(\sigma_0\sigma_1\sigma_2)=0, & \\
	\\
\; [O_4]\equiv\{\sigma_0\sigma_1\} \rightarrow & 
	x^4 + d x^3 + e x^2 + f x + 1 & \rightarrow x^4 + x^3 + x^2 + x + 1, \\
 & d = Tr(\sigma_0\sigma_1) = 1, & \\
 & e = Tr(\sigma_0\sigma_1) = 1, & \\
 & f = Tr(\sigma_0\sigma_1) = 1, & \\
	\\
\; [O_2]\equiv\{\sigma_0\sigma_2\} \rightarrow & 
	x^2 + g x + 1 & \rightarrow x^2 + x + 1, \\
 & g = Tr(\sigma_0\sigma_2) = 1, & \\
	\\
\; [O_1]\equiv\{\sigma_0\sigma_1\sigma_2\sigma_3\} \rightarrow & 
	 x + 1 & \rightarrow x + 1, \\
	\\
\; [O_1]\equiv\{0\} \rightarrow & 
	 x  & \rightarrow x, 
\end{eqnarray*}
where $\{\sigma_i\cdots\sigma_k\}$ is representing all the $GF(2^4)$ elements 
that correspond to the orbital [$O_m$].

The step to find the $+$-representation of every $a\in GF(2^4)$ is the following. \newline
Let us take $\sigma_0$ with associated I.P.: $x^4 + x^3 + 1$, then:
\begin{eqnarray*}
\sigma_0^4 + \sigma_0^3 + 1  & = & 0, \\
\sigma_0^4 = \sigma_2 & \rightarrow & (\sigma_0^3)_+ = 1 + (\sigma_2)_+ = 1011.
\end{eqnarray*}
Taking the shifts of $1011$, we find the $+$-representation for all the elements 
of the orbital $\{\sigma_0^3\}\equiv\{\sigma_0\sigma_1\}$. The distributive
property determines the $+$-representation of the orbit $\{\sigma_0\sigma_1\sigma_2\}$,
with the final result shown in table~\ref{tabla5}. \newline
Let us observe that this construction of $GF(2^4)$ is unique.\newline

 {\bf n=5}:
We repeat the same process as one of the case $n=4$.
Let us suppose $GF(2^5)$  generated by the primitive normal
basis $B_{\sigma 5}=\{\sigma_0,\sigma_1,\sigma_2,\sigma_3,\sigma_4\}$.
Hence $\sigma_{i}^2=\sigma_{(i+1)\bmod 2}$. The field's elements  
$\{0,\sigma_0,\sigma_0^2,\sigma_0^3,\cdots,\sigma_0^{30},\sigma_0^{31}=1\}$
are generated by $\sigma_0$. 
If an element $a\in GF(2^5)$ verifies $a^{2^m}=a$, then $m\mid 5$ and 
the $m$ powers of $a$, $\{a,a^2,a^{2^2},\cdots,a^{2^{m-1}}\}$,
are the $m$ roots of an I.P. of degree $m$ over $GF(2)$.
The representative different subsets of this kind in $GF(2^5)$ are:
\begin{center}
\begin{tabular}{|l|l|} \hline
 $a\in GF(2^5)$      & $m$           \\ \hline\hline 
 $0 $         & $1$            \\
 $\sigma_0$   & $5$ 	   \\
 $\sigma_0\sigma_1$ & $5$      \\
 $\sigma_0\sigma_2$  & $5$     \\
 $\sigma_0\sigma_1\sigma_2$ & $5$          \\
 $\sigma_0\sigma_1\sigma_3$  & $5$         \\
 $\sigma_0\sigma_1\sigma_2\sigma_3$ & $5$  \\
 $\sigma_0\sigma_1\sigma_2\sigma_3\sigma_4$ & $1$  \\ \hline
\end{tabular}
\end{center}
The  different orbitals $[O_m]$ of period $m$, with $m\mid 5$,
in ${\cal B}_5$ are given by the representative orbits:
\begin{center}
\begin{tabular}{|l|l|l|} \hline
$[O_m]$  & $Tr(O_m)$  & m               \\ \hline\hline 
$00000$ &      0     & 1  \\
$00001$ &      1     & 5  \\
$00011$ &      0     & 5  \\  
$00101$ &      0     & 5  \\
$00111$ &      1     & 5  \\
$01011$ &      1     & 5  \\
$01111$ &      0     & 5  \\
$11111$ &      1     & 1  \\   \hline
\end{tabular}
\end{center}
Thus, $N_T^{(5)}=N_1+N_5=8$ and $N_{T1}^{(5)}=4$.\newline
If $\{\sigma_0,\sigma_1,\sigma_2,\sigma_3,\sigma_4\}$ are taken as the roots of an I.P. of degree 5
the equations derived from Newton's formulas (where $\theta_k=0$ for $k>5$) for the different traces
can be established. The variables $\theta_k$ and $s_k$ are expressed as follows: 
\begin{eqnarray*}
\theta_1  & = & \sigma_0 + \sigma_1 + \sigma_2 + \sigma_3  \\
	& = & Tr(\sigma_0) = 1,\\
\theta_2  & = & \sigma_0\sigma_1 + \sigma_1\sigma_2 + \sigma_2\sigma_3 + \sigma_3\sigma_4 +\sigma_4\sigma_0  \\
	&   & +\sigma_0\sigma_2 + \sigma_1\sigma_3 + \sigma_2\sigma_4 + \sigma_3\sigma_0 + \sigma_4\sigma_1  \\
	& = & Tr(\sigma_0\sigma_1) + Tr(\sigma_0\sigma_2), \\
\theta_3 & = & \sigma_0\sigma_1\sigma_2 + \sigma_1\sigma_2\sigma_3 + 
	\sigma_2\sigma_3\sigma_4 +  \sigma_3\sigma_4\sigma_0 + \sigma_4\sigma_0\sigma_1 \\
	& & +\sigma_0\sigma_1\sigma_3 + \sigma_1\sigma_2\sigma_4 + 
	\sigma_2\sigma_3\sigma_0 +  \sigma_3\sigma_4\sigma_1 + \sigma_4\sigma_0\sigma_2 \\
	& = & Tr(\sigma_0\sigma_1\sigma_2) + Tr(\sigma_0\sigma_1\sigma_3), \\
\theta_4 & = & \sigma_0\sigma_1\sigma_2\sigma_3 +\sigma_1\sigma_2\sigma_3\sigma_4 +\sigma_2\sigma_3\sigma_4\sigma_0 \\
	 & & +\sigma_3\sigma_4\sigma_0\sigma_1+\sigma_4\sigma_0\sigma_1\sigma_2 \\
	& = & Tr(\sigma_0\sigma_1\sigma_2\sigma_3), \\
\theta_5 & = & Tr(\sigma_0\sigma_1\sigma_2\sigma_3\sigma_4)  = 1,\\
s_k	& = & \sigma_0^k + \sigma_1^k + \sigma_2^k + \sigma_3^k + \sigma_4^k \\
	& = & \sum_i Tr(\sigma^i) \;\; with\;\; k=1,2,\cdots,31.
\end{eqnarray*}
The independent Newton's formulas from the system~(\ref{eqsk}) are:
\begin{eqnarray*}
Tr(\sigma_0\sigma_2) + Tr(\sigma_0\sigma_1\sigma_2) + Tr(\sigma_0\sigma_1\sigma_3) & = & 1,  \\
Tr(\sigma_0\sigma_1\sigma_3) + Tr(\sigma_0\sigma_1\sigma_2\sigma_3)  & = & 0, \\
Tr(\sigma_0\sigma_1) + Tr(\sigma_0\sigma_1\sigma_2) & = & 0, \\
Tr(\sigma_0\sigma_2)(1+Tr(\sigma_0\sigma_1\sigma_3)) & = & 0, \\
Tr(\sigma_0\sigma_2)(1+Tr(\sigma_0\sigma_1)) & = & 0. \\
\end{eqnarray*}
One solution compatible with the constraints is:
\begin{eqnarray*}
Tr(\sigma_0) & = & 1, \\
Tr(\sigma_0\sigma_1) & = & 0, \\
Tr(\sigma_0\sigma_2) & = & 0, \\
Tr(\sigma_0\sigma_1\sigma_2) & = & 0, \\
Tr(\sigma_0\sigma_1\sigma_3) & = & 1, \\
Tr(\sigma_0\sigma_1\sigma_2\sigma_3) & = & 1, \\
Tr(\sigma_0\sigma_1\sigma_2\sigma_3\sigma_4) & = & 1. \\
\end{eqnarray*}
Finally, the I.P. are calculated from this solution:
\begin{eqnarray*}
\; [O_5]\equiv\{\sigma_0\} \rightarrow & 
	x^5 + a x^4 + b x^3 + c x^2 +  d x + 1 & \rightarrow x^5 + x^4 + x^2 \\
 	& & 							\;\;\; + x + 1, \\
\; [O_5]\equiv\{\sigma_0^{-1}\equiv \sigma_0\sigma_1\sigma_2\sigma_3\} \rightarrow & 
	x^5 + d x^4 + c x^3 + b x^2 +  a x + 1	 & \rightarrow x^5 + x^4 + x^3 \\
	& &							\;\;\; + x + 1,  \\
 & a = Tr(\sigma_0) = 1, & \\
 & b = Tr(\sigma_0\sigma_1) + Tr(\sigma_0\sigma_2) = 0, & \\
 & c = Tr(\sigma_0\sigma_1\sigma_2) + Tr(\sigma_0\sigma_1\sigma_3) = 1, & \\
 & d = Tr(\sigma_0\sigma_1\sigma_2\sigma_3) = 1, & \\
       \\
\; [O_5]\equiv\{\sigma_0\sigma_1\} \rightarrow & 
	x^5 + e x^4 + f x^3 + g x^2 +  h x + 1 & \rightarrow x^5 + x^3 + 1, \\
\; [O_5]\equiv\{(\sigma_0\sigma_1)^{-1}\equiv \sigma_0\sigma_1\sigma_2\} \rightarrow & 
	x^5 + h x^4 + g x^3 + f x^2 +  e x + 1	 & \rightarrow x^5 + x^2 + 1,  \\
 & e = Tr(\sigma_0\sigma_1) = 0, & \\
 & f = Tr(\sigma_0\sigma_2) + Tr(\sigma_0\sigma_1\sigma_2\sigma_3) = 1, & \\
 & g = Tr(\sigma_0\sigma_1\sigma_3) + Tr(\sigma_0) = 1, & \\
 & h = Tr(\sigma_0\sigma_1\sigma_2) = 0, & \\
       \\
\; [O_5]\equiv\{\sigma_0\sigma_2\} \rightarrow & 
	x^5 + j x^4 + k x^3 + l x^2 +  m x + 1 & \rightarrow x^5 + x^3 + x^2 \\
	& &							\;\;\; + x + 1, \\
\; [O_5]\equiv\{(\sigma_0\sigma_2)^{-1}\equiv \sigma_0\sigma_1\sigma_3\} \rightarrow & 
	x^5 + m x^4 + l x^3 + k x^2 +  j x + 1	 & \rightarrow x^5 + x^4 + x^3 \\
	& &							\;\;\; + x^2 + 1,  \\
 & j = Tr(\sigma_0\sigma_2) = 0, & \\
 & k = Tr(\sigma_0\sigma_1\sigma_2\sigma_3) + Tr(\sigma_0\sigma_1\sigma_2) = 1, & \\
 & l = Tr(\sigma_0) + Tr(\sigma_0\sigma_1) = 1, & \\
 & m = Tr(\sigma_0\sigma_1\sigma_3) = 1, & \\
	\\
\; [O_1]\equiv\{\sigma_0\sigma_1\sigma_2\sigma_3\sigma_4\} \rightarrow & 
	 x + 1 & \rightarrow x + 1, \\
	\\
\; [O_1]\equiv\{0\} \rightarrow & 
	 x  & \rightarrow x, 
\end{eqnarray*}
where $\{\sigma_i\cdots\sigma_k\}$ is representing all the $GF(2^5)$ elements that
correspond to the orbital [$O_m$].

The steps to find the $+$-representation of every $a\in GF(2^5)$ are the following. \newline
Let us take $\sigma_0$ with associated I.P.: $x^5 + x^4 + x^2 + x + 1$, then:
\begin{eqnarray*}
\sigma_0^5 + \sigma_0^4 + \sigma_0^2 + \sigma_0 + 1  & = & 0, \\
\sigma_0^2 = \sigma_1\; ;\;\sigma_0^4 = \sigma_2 & \rightarrow & 
(\sigma_0^5)_+ = 1 + (\sigma_0)_+ +(\sigma_1)_+ +(\sigma_2)_+ = 11000.
\end{eqnarray*}
The shifts of $11000$ are the $+$-representation for all the elements 
of the orbital $\{\sigma_0^5\}\equiv\{\sigma_0\sigma_2\}$. \newline
If we now take $\sigma_0^3$ with associated I.P.: $x^5 + x^3 + 1$, then:
\begin{eqnarray*}
(\sigma_0^3)^5 + (\sigma_0^3)^3 + 1  & = & 0, \\
(\sigma_0^9)_+ = 00110  & \rightarrow & 
(\sigma_0^{15})_+ = 1 + (\sigma_0^9)_+ = 11001.
\end{eqnarray*}
The shifts of $11001$ are the $+$-representation for all the elements 
of the orbital $\{\sigma_0^{15}\}\equiv\{\sigma_0\sigma_1\sigma_2\sigma_3\}$. \newline
Proceeding now with $\sigma_0^7$ that has an associated I.P.: $x^5 + x^2 + 1$, then:
\begin{eqnarray*}
(\sigma_0^7)^5 + (\sigma_0^7)^2 + 1  & = & 0, \\
\sigma_0^{35}= \sigma_0^4  & \rightarrow & 
(\sigma_0^{14})_+ = 1 + (\sigma_0^4)_+ = 11011.
\end{eqnarray*}
The shifts of $11011$ are the $+$-representation for all the elements 
of the orbital $\{\sigma_0^{7}\}\equiv\{\sigma_0\sigma_1\sigma_2\}$. \newline
Taking $\sigma_0^{15}$ with associated I.P.: $x^5 + x^4 + x^3 + x + 1$, then:
\begin{eqnarray*}
(\sigma_0^{15})^5 + (\sigma_0^{15})^4 + (\sigma_0^{15})^3 + \sigma_0^{15} + 1  & = & 0, \\
\sigma_0^{60}=00111\; ;\;\sigma_0^{45}=11011\;   & \rightarrow & 
(\sigma_0^{75})_+ = (\sigma_0^{13})_+ = \\
 & & 1 + (\sigma_0^{15})_+ + (\sigma_0^{45})_+ + (\sigma_0^{60})_+ = 11010.
\end{eqnarray*}
The shifts of $11010$ are the $+$-representation for all the elements 
of the orbital $\{\sigma_0^{11}\}\equiv\{\sigma_0\sigma_1\sigma_3\}$. \newline
Finally, the $+$-representation for the elements 
of the orbital $\{\sigma_0^{3}\}\equiv\{\sigma_0\sigma_1\}$ are the shifts of $01001$.
Other solutions for the Newton's formulas can allow different natural representations 
of the field $GF(2^5)$ in ${\cal B}_n$. Table~\ref{tabla6} is one of them.

\section{Conclusion}

A possible method to build up the irreducible polynomials of degree $n$
over the field $GF(2)$ has been exposed in this paper. The strategy
exploits the properties and the connection between the splitting field $GF(2^n)$
and the space of $n$-binary sequences ${\cal B}_n$. 

The choice of a primitive normal basis allows us to perform a natural 
construction of $GF(2^n)$  in ${\cal B}_n$:
(i) Two natural operations $(+,*)$ are defined on ${\cal B}_n$ in order 
to obtain a representation of the splitting field $GF(2^n)$;
(ii) A system of equations for the different traces of the field is derived
 from the fundamental theorem on symmetrical polynomials;
(iii) Different constraints, that have been established, limit the number 
of possible solutions of these equations;
(iv) Any of these solutions gives us the irreducible polynomials of degree $m$, 
 with $m\mid n$;
(v) The field $({\cal B}_n,+,*)$ is finally built up with the help of 
the irreducible polynomials.

We want to remark the direct relationship that seems to exist between the irreducible polynomials
over $GF(2)$ and the periodic binary orbits, in such a way that this is a one-to-one application. 
Following this ideas and under a similar scheme, this method could be applied to find 
the irreducible polynomials of degree $n$ over a general finite field $GF(q)$ by 
using the connection between the properties of 
the splitting field $GF(q^n)$ and the space of $n$-q-ary sequences.

{\bf Acknowledgements:}
This work was performed during a postdoctoral stage at the Ecole Normale Sup\'erieure of Paris
in 1995. I thank J-M. Couveignes (Paris) and J. Cabeza (Zaragoza)
for useful discussions and references on this subject of finite fields,
and the Gouvernement Fran\c{c}ais for a research grant.\newline
\underline{Note}: This manuscript should be understood as the approach of 
a 'newbie', plenty of questions, to a subject very far from its common working topics. 

\newpage

\appendix{}
\begin{center} {\bf Appendix: Calculation of $N_{T}^{(n)}$ and $N_{T1}^{(n)}$} \end{center}

An algorithm to calculate $N_{T}^{(n)}$ and $N_{T1}^{(n)}$ is proposed in this appendix. 
The process to find the different orbitals in ${\cal B}_n$ involves the following steps:

(1) The decimal value of the different $n$-digit orbitals is stored in an indexed 
integer variable $O(i)$, then $0<i<N_{T}^{(n)}$.
The initial conditions are: $O(0)=0$, $O(1)=1$. 
Another integer variable $O_p$ is defined and initialized as $O_p=1$.

(2) The operational loop to find $O(k+1)$ when all orbitals $O(m)$ are known for $m\leq k$ is the following:

\indent $\;\hspace{3cm} (**)\; O_p = O_p+1$ \newline
\indent $\hspace{4cm} do\;\; i=0,n-1$ \newline
\indent $\hspace{4.5cm} \overline{O_p} = O_p\cdot 2^i \bmod (2^n-1)$ \newline
\indent $\;\hspace{4.5cm} do\;\; \;\;j=1,k$ \newline
\indent $\;\hspace{5cm} if\;(\overline{O_p}.eq.O(j))\;\; goto\; (**)$ \newline
\indent $\;\hspace{3cm} \hspace{1.5cm} end\; do$ \newline
\indent $\;\hspace{4cm} end\; do$ \newline
\indent $\;\hspace{4cm} O(k+1) = O_p$ \newline
\indent $\;\hspace{4cm} k = k+1 $\newline 
\indent $\;\hspace{4cm} goto\; (**)$\newline

(3) This is repeated for all sequences of ${\cal B}_n$ while $O_p\leq (2^n-1)$.
The maximum index $k$ from the step (2) is $k_{max} = N_T^{(n)}-1$.

(4) The period ($m$) of each orbital $O(i)$ is calculated:
\begin{eqnarray*}
per(O(i)) = m  & \Leftrightarrow & m\; is\;the\; smallest\; integer\;\;1\leq m\leq n-1 \\
 & & \rightarrow  O(i) = O(i)\cdot 2^m \bmod (2^n-1)    
\end{eqnarray*}

(5) Each orbital is converted to its binary expression and its trace is calculated:
\begin{eqnarray*}
 O_b(i) & = & binary(O(i))  \\
Tr(O(i))\equiv Tr(O_b(i)) & = & O_b(i) + shift(O_b(i)) + shift^2(O_b(i))+\cdots + shift^m(O_b(i)) 
\end{eqnarray*}

(6) The number of traces equal to $1$ when $i$ ranges from $1$ to $N_{T}^{(n)}$ is $N_{T1}^{(n)}$.

\newpage
\begin{center} Table Captions \end{center}

{\bf Table 1.} Functions over the natural numbers $\cal N$, that appear 
in this work, are calculated for the first $n\in\cal N$.
$N_n$ is the number of irreducible polynomials of degree $n$ over $GF(2)$.
$N_T^{(n)}=\sum_m N_m$ with $m|n$ represents the total number 
of orbitals in ${\cal B}_n$.
$N_{T1}^{(n)}$ and $N_{T0}^{(n)}$ are the number of orbitals in ${\cal B}_n$
with trace $1$ or $0$, respectively.

{\bf Table 2,3,4,5,6.} Natural representation of the finite field $GF(2^n)$ 
for the cases $n=1,2,3,4,5$ in the space of binary sequences ${\cal B}_n$,
where the two internal operations are performed in the following way: 
(i) {\it addition}: $(a_++b_+)_+=a_++b_+$  ; 
(ii){\it multiplication}: $(a*b)_*=a_**b_*$. 
The set (${\cal B}_n,+,*$) is a representation of the finite field $GF(2^n)$. 
The $m$ orbits that compose each orbital $[O_m]$, with $m\mid n$, are the $m$ roots
of an irreducible polynomial (I.P.) of degree $m$ over $GF(2)$. 

\newpage

\begin{table}[htb]
\begin{center}
\begin{tabular}{|c|c|c|c|c|} \hline
 $n$   &  $N_n$ & $N_T^{(n)}$  & $N_{T1}^{(n)}$ & $N_{T0}^{(n)}$  \\ \hline \hline
 $1$   &   $2$  &  $2$     &    $1$    &  $1$     \\ \hline
 $2$   &   $1$  &  $3$     &    $2$    &  $1$     \\ \hline
 $3$   &   $2$  &  $4$     &    $2$    &  $2$     \\ \hline
 $4$   &   $3$  &  $6$     &    $4$    &  $2$     \\ \hline
 $5$   &   $6$  &  $8$     &    $4$    &  $4$     \\ \hline
 $6$   &   $9$  &  $14$    &    $8$    &  $6$     \\ \hline
 $7$   &   $18$ &  $20$    &    $10$    &  $10$     \\ \hline
\end{tabular}
\caption{}
\label{tabla1}
\end{center}
\end{table}

\begin{table}[htb]
\begin{center}
\begin{tabular}{|l|l|c|c|c|c|} \hline
\multicolumn{6}{|c|}{$GF(2)$} \\ \hline\hline
\multicolumn{2}{|c|}{element: $a$}  & $a_*$  & $a_+$ & $Trace$ & $I.P.$  \\ \hline
$\sigma_0$    &    $\sigma_0$     &   $1$  & $1$   &  $1$    &       $x+1$               \\ \hline
       $0$    &         $0$     &   $0$   & $0$   &  $0$    &       $x$           \\ \hline
\end{tabular}
\caption{}
\label{tabla2}
\end{center}
\end{table}

\begin{table}[htb]
\begin{center}
\begin{tabular}{|l|l|l|c|c|c|c|} \hline
\multicolumn{7}{|c|}{$GF(2^2)$} \\ \hline\hline
\multicolumn{3}{|c|}{element: $a$}  & $a_*$  & $a_+$ & $Trace$ & $I.P.$  \\ \hline
$\sigma_0$    & $\sigma_0$  & $\sigma_0$     &  $01$  & $01$  &  $1$    &       $x^2+x+1$         \\ 
$\sigma_0^2$  & $\sigma_1$  & $\sigma_1$     &  $10$  & $10$  &         &                         \\ \hline
$\sigma_0^3$ & $\sigma_0+\sigma_1$ & $\sigma_0\sigma_1$ &  $11$  & $11$  &  $1$    &       $x+1$         \\ \hline
       $0$    &      $0$   &     $0$     &    $00$    & $00$  &  $0$    &       $x$               \\ \hline
\end{tabular}
\caption{}
\label{tabla3}
\end{center}
\end{table}

\begin{table}[htb]
\begin{center}
\begin{tabular}{|l|l|l|c|c|c|c|} \hline
\multicolumn{7}{|c|}{$GF(2^3)$} \\ \hline\hline
\multicolumn{3}{|c|}{element: $a$}  & $a_*$  & $a_+$ & $Trace$ & $I.P.$  \\ \hline
$\sigma_0$      & $\sigma_0$ & $\sigma_1$    &   $001$  & $001$   &         &                   \\
$\sigma_0^2$    & $\sigma_1$ & $\sigma_2$    &   $010$  & $010$   &  $1$    &   $x^3+x^2+1$     \\ 
$\sigma_0^4$    & $\sigma_2$ & $\sigma_3$    &   $100$  & $100$   &         &                   \\ \hline
$\sigma_0^3$    & $\sigma_0+\sigma_2$ & $\sigma_0\sigma_1$  &   $011$  & $101$   &      &              \\ 
$\sigma_0^6$    & $\sigma_1+\sigma_0$ & $\sigma_1\sigma_2$  &   $110$  & $011$   & $0$  &   $x^3+x+1$  \\ 
$\sigma_0^5$    & $\sigma_2+\sigma_1$ & $\sigma_2\sigma_0$  &   $101$  & $110$   &      &              \\ \hline
$\sigma_0^7$  & $\sigma_0+\sigma_1+\sigma_2$ & $\sigma_0\sigma_1\sigma_2$ & $111$ & $111$ & $1$ & $x+1$  \\ \hline
     $0$        &     $0$   &    $0$           &    $000$    & $000$   &  $0$ &       $x$    \\ \hline
\end{tabular}
\caption{}
\label{tabla4}
\end{center}
\end{table}

\begin{table}[htb]
\begin{center}
\begin{tabular}{|l|l|l|c|c|c|c|} \hline
\multicolumn{7}{|c|}{$GF(2^4)$} \\ \hline\hline
\multicolumn{3}{|c|}{element: $a$}  & $a_*$  & $a_+$ & $Trace$ & $I.P.$    \\ \hline
$\sigma_0$      & $\sigma_0$ & $\sigma_0$    &   $0001$  & $0001$   &         &                   \\ 
$\sigma_0^2$    & $\sigma_1$ & $\sigma_1$    &   $0010$  & $0010$   &  $1$    &   $x^4+x^3+1$     \\
$\sigma_0^4$    & $\sigma_2$ & $\sigma_2$    &   $0100$  & $0100$   &         &                   \\ 
$\sigma_0^8$    & $\sigma_3$ & $\sigma_3$    &   $1000$  & $1000$   &         &                   \\ \hline
$\sigma_0^3$ & $\sigma_0+\sigma_1+\sigma_3$ &  $\sigma_0\sigma_1$ 
&   $0011$  & $1011$   &      &                   \\
$\sigma_0^6$    &  $\sigma_1+\sigma_2+\sigma_0$  & $\sigma_1\sigma_2$     
&   $0110$  & $0111$   & $1$  &  $x^4+x^3+x^2+x+1$ \\ 
$\sigma_0^{12}$ & $\sigma_2+\sigma_3+\sigma_1$ & $\sigma_2\sigma_3$     
&   $1100$  & $1110$   &      &                    \\
$\sigma_0^9$    & $\sigma_3+\sigma_0+\sigma_2$ &   $\sigma_3\sigma_0$     
&   $1001$  & $1101$   &      &                    \\ \hline
$\sigma_0^5$    &  $\sigma_0+\sigma_2$ & $\sigma_0\sigma_2$     
&   $0101$  & $0101$   & $1$  & $x^2+x+1$          \\ 
$\sigma_0^{10}$ &  $\sigma_1+\sigma_3$ & $\sigma_1\sigma_3$     
&   $1010$  & $1010$   &      &                    \\ \hline
$\sigma_0^{7}$  &  $\sigma_2+\sigma_3$ & $\sigma_0\sigma_1\sigma_2$     
&   $0111$  & $1100$   &      &                    \\ 
$\sigma_0^{14}$ &   $\sigma_3+\sigma_0$ & $\sigma_1\sigma_2\sigma_3$     
&   $1110$  & $1001$   & $0$  &   $x^4+x+1$        \\ 
$\sigma_0^{13}$ &    $\sigma_0+\sigma_1$ & $\sigma_2\sigma_3\sigma_0$     
&   $1101$  & $0011$   &      &                    \\ 
$\sigma_0^{11}$ &   $\sigma_1+\sigma_2$ & $\sigma_3\sigma_0\sigma_1$     
&   $1011$  & $0110$   &      &                    \\ \hline
$\sigma_0^{15}$ &  $\sigma_0+\sigma_1+\sigma_2+\sigma_3$ &  $\sigma_0\sigma_1\sigma_2\sigma_3$     
&   $1111$  & $1111$   &  $1$ &  $x+1$     \\ \hline
     $0$        &           $0$   &    $0$    &     $0000$ & $0000$   &  $0$ &  $x$       \\ \hline
\end{tabular}
\caption{}
\label{tabla5}
\end{center}
\end{table}

\begin{table}[htb]
\begin{center}
\begin{tabular}{|l|l|l|c|c|c|c|} \hline
\multicolumn{7}{|c|}{$GF(2^5)$} \\ \hline\hline
\multicolumn{3}{|c|}{element: $a$}  & $a_*$  & $a_+$ & $Trace$ & $I.P.$  \\ \hline
$\sigma_0$      & $\sigma_0$  & $\sigma_0$  &   $00001$  & $00001$   &         &                       \\ 
$\sigma_0^2$    & $\sigma_1$  & $\sigma_0$  &   $00010$  & $00010$   &         &  $x^5+x^4+x^2$        \\ 
$\sigma_0^4$    & $\sigma_2$  & $\sigma_2$  &   $00100$  & $00100$   &  $1$    &  $+ x+1$    \\ 
$\sigma_0^8$    & $\sigma_3$  & $\sigma_3$  &   $01000$  & $01000$   &         &                       \\ 
$\sigma_0^{16}$ & $\sigma_4$  & $\sigma_4$  &   $10000$  & $10000$   &         &                       \\ \hline
$\sigma_0^3$    & $\sigma_0+\sigma_3$ &  $\sigma_0\sigma_1$ &   $00011$  & $01001$   &      &                \\ 
$\sigma_0^6$    & $\sigma_1+\sigma_4$ &  $\sigma_1\sigma_2$ &   $00110$  & $10010$   &      &                \\ 
$\sigma_0^{12}$ & $\sigma_2+\sigma_0$ &  $\sigma_2\sigma_3$ &   $01100$  & $00101$   &  $0$ &   $x^5+x^3+1$  \\
$\sigma_0^{24}$ & $\sigma_3+\sigma_1$ &  $\sigma_3\sigma_4$ &   $11000$  & $01010$   &      &                \\
$\sigma_0^{17}$ & $\sigma_4+\sigma_2$ &  $\sigma_4\sigma_0$ &   $10001$  & $10100$   &      &                \\ \hline
$\sigma_0^5$     & $\sigma_3+\sigma_4$ & $\sigma_0\sigma_2$  &   $00101$  & $11000$   &      &               \\
$\sigma_0^{10}$  & $\sigma_4+\sigma_0$ & $\sigma_1\sigma_3$  &   $01010$  & $10001$   &      &   $x^5+x^3+x^2$   \\
$\sigma_0^{20}$  & $\sigma_0+\sigma_1$ & $\sigma_2\sigma_4$  &   $10100$  & $00011$   &  $0$ &   $+ x+1$          \\
$\sigma_0^{9}$   & $\sigma_1+\sigma_2$ & $\sigma_3\sigma_0$  &   $01001$  & $00110$   &      &               \\
$\sigma_0^{18}$  & $\sigma_2+\sigma_3$ & $\sigma_4\sigma_1$  &   $10010$  & $01100$   &      &               \\ \hline
$\sigma_0^{7}$  & $\sigma_0+\sigma_2+\sigma_3+\sigma_4$ &  $\sigma_0\sigma_1\sigma_2$     
&   $00111$  & $11101$   &      &                    \\
$\sigma_0^{14}$ & $\sigma_1+\sigma_3+\sigma_4+\sigma_0$ &  $\sigma_1\sigma_2\sigma_3$     
&   $01110$  & $11011$   &      &                    \\ 
$\sigma_0^{28}$ & $\sigma_2+\sigma_4+\sigma_0+\sigma_1$ &  $\sigma_2\sigma_3\sigma_4$     
&   $11100$  & $10111$   & $0$  &   $x^5+x^2+1$      \\ 
$\sigma_0^{25}$ & $\sigma_3+\sigma_0+\sigma_1+\sigma_2$ & $ \sigma_3\sigma_4\sigma_0$     
&   $11001$  & $01111$   &      &                    \\ 
$\sigma_0^{19}$ & $\sigma_4+\sigma_1+\sigma_2+\sigma_3$ & $\sigma_4\sigma_0\sigma_1$     
&   $10011$  & $11110$   &      &                    \\                                       \hline
$\sigma_0^{11}$ &  $\sigma_1+\sigma_2+\sigma_4$ & $\sigma_0\sigma_1\sigma_3$ 
&   $01011$  & $10110$   &      &                           \\ 
$\sigma_0^{22}$ &  $\sigma_2+\sigma_3+\sigma_0$ & $\sigma_1\sigma_2\sigma_4$  
&   $10110$  & $01101$   &      &   $x^5+x^4+x^3$              \\ 
$\sigma_0^{13}$ &  $\sigma_3+\sigma_4+\sigma_1$ & $\sigma_2\sigma_3\sigma_0$  
&   $01101$  & $11010$   & $1$  &   $+ x^2+1$                   \\
$\sigma_0^{26}$ & $\sigma_4+\sigma_0+\sigma_2$ & $\sigma_3\sigma_4\sigma_1$  
&   $11010$  & $10101$   &      &                           \\ 
$\sigma_0^{21}$ & $\sigma_0+\sigma_1+\sigma_3$ &  $\sigma_4\sigma_0\sigma_2$  
&   $10101$  & $01011$   &      &                           \\                                 \hline
$\sigma_0^{15}$ &   $\sigma_0+\sigma_3+\sigma_4$ &   $\sigma_0\sigma_1\sigma_2\sigma_3$ 
&   $01111$  & $11001$   &      &                   \\
$\sigma_0^{30}$ &   $\sigma_1+\sigma_4+\sigma_0$ &  $\sigma_1\sigma_2\sigma_3\sigma_4$  
&   $11110$  & $10011$   &      &    $x^5+x^4+x^3$       \\ 
$\sigma_0^{29}$ &   $\sigma_2+\sigma_0+\sigma_1$ &  $\sigma_2\sigma_3\sigma_4\sigma_0$  
&   $11101$  & $00111$   & $1$  & $+ x+1$  \\ 
$\sigma_0^{27}$ &   $\sigma_3+\sigma_1+\sigma_2$ &  $\sigma_3\sigma_4\sigma_0\sigma_1$  
&   $11011$  & $01110$   &      &                   \\
$\sigma_0^{23}$ &   $\sigma_4+\sigma_2+\sigma_3$ &  $\sigma_4\sigma_0\sigma_1\sigma_2$  
&   $10111$  & $11100$   &      &                   \\                                         \hline
$\sigma_0^{31}$ & $\sigma_0+\sigma_1+\sigma_2+\sigma_3+\sigma_4$ & $\sigma_0\sigma_1\sigma_2\sigma_3\sigma_4$ 
&   $11111$  & $11111$   & $1$  &  $x+1$     \\ \hline
     $0$        &     $0$  &      $0$              &    $00000$    & $00000$   &  $0$ &       $x$    \\ \hline
\end{tabular}
\caption{}
\label{tabla6}
\end{center}
\end{table}

\end{document}